# Data Engineering Patterns for Cross-System Reconciliation in Regulated Enterprises: Architecture, Anomaly Detection, and Governance

Zhijun Qiu

*Independent Practitioner, Enterprise Data Engineering and Governance*

qrit110@gmail.com

ORCID: 0009-0006-6335-6831

**ABSTRACT**

Regulated enterprises in the United States—banks, telecommunications providers, large technology companies—operate across heterogeneous systems that were rarely designed to interoperate. ERP platforms, billing engines, supply chain tools, and financial reporting infrastructure coexist within the same organization, but they do not talk to each other well. The resulting fragmentation produces familiar problems: transactions recorded in one system but unreconciled in another, asset inventories drifting from their systems of record, and audit-readiness that depends on manual effort. The PCAOB's 2024 inspection cycle put a number on the consequences: a 39% aggregate Part I.A deficiency rate across all inspected firms [26]. This paper introduces the GERA Framework (Governed Enterprise Reconciliation Architecture)—a vendor-neutral, four-layer data architecture that integrates deterministic cross-system reconciliation, statistical anomaly detection (baseline Z-Score with robust alternatives), governed semantic standardization, and NIST CSF 2.0-aligned security controls into a single methodology. The architecture spans four layers (ingestion, staging, core models, and semantic serving), following the multi-layer pattern now common in modern data platforms [2]. The patterns are demonstrated through U.S. broadband operations—where billing reconciliation, inventory aging, and governance are tightly coupled—and draw on the author's implementation experience across three regulated enterprise environments: a regional bank, a national broadband provider, and a Fortune 500 technology company's central finance organization. This is a practitioner reference—an architectural framework paper documenting field-tested patterns—not a controlled experiment or benchmark study. No proprietary systems, datasets, or internal implementations are disclosed.

## 1. INTRODUCTION

Anyone who has spent time inside a regulated enterprise's data environment knows the pattern. Financial close depends on a spreadsheet someone emails on the third business day. Asset counts require a physical walk because the system of record drifted months ago. A reconciliation dispute takes weeks to resolve because the ERP and the downstream reporting platform use different identifiers. The specifics vary by industry—in banking, general ledger transactions must reconcile across sub-ledger systems under SOX mandates; in telecommunications, provisioning events must align with billing records; in enterprise technology, financial reporting pipelines must satisfy both internal controls and external audit requirements—but the underlying failure mode is the same. The PCAOB's 2024 inspection cycle reported a 39% aggregate Part I.A deficiency rate across all inspected firms [26]. Cross-system data accuracy in financial reporting is not a niche problem. This paper is a practitioner reference, not an empirical study. It documents architectural patterns drawn from field experience and presents them as a reproducible framework; it does not report controlled experiments or measured performance outcomes.

These are not edge cases. For regulated enterprises across sectors, fragmented operational data is a structural condition. The technology stacks that run enterprise operations—ERP platforms, billing and revenue systems, procurement tools, financial reporting infrastructure—were typically acquired at different times, from different vendors, to solve different problems. The result is a patchwork of data sources with no common keys, no shared definitions, and no unified governance model. This paper introduces the GERA Framework (Governed Enterprise Reconciliation Architecture)—a four-layer data architecture that integrates deterministic cross-system reconciliation, statistical anomaly detection, governed semantic standardization, and NIST CSF 2.0-aligned security controls. U.S. broadband operations provide a particularly sharp illustration, where Internet Service Providers (ISPs) face these problems across Operations Support Systems (OSS/BSS), procurement tools, contractor portals, and network management systems [4].

Three bottlenecks recur across this landscape. The examples below are drawn from broadband, but anyone who has worked in banking data reconciliation or enterprise financial reporting will recognize the pattern.

### 1.1 Inventory Invisibility

Broadband construction is materials-intensive. Fiber cable, conduit, splice closures, optical network terminals, and network interface devices flow through a chain that may include the ISP's own warehouses, third-party logistics providers, and field contractors. Tracking this inventory means reconciling purchase orders, receiving records, issuance documents, and installation confirmations across organizational boundaries—often with no shared system of record between parties. When that reconciliation breaks down, or never existed, providers lose visibility into what they own, where it sits, and how long it has been sitting there. The consequences can show up as over-purchasing that ties up working capital, shortages that stall construction crews, and obsolete stock that quietly accumulates.

### 1.2 Revenue Leakage and Billing Disputes

A broadband subscriber's lifecycle touches multiple systems: a CRM or order management platform captures the sale, a provisioning system activates the service, a billing platform generates invoices, and a payments system records settlements. Each may represent the same subscriber differently—different account identifiers, different timestamps, different event granularity. When these systems fall out of alignment, the results are predictable: services activated but never invoiced, duplicate charges, credits applied without corresponding service events, and reconciliation disputes that consume analyst time for weeks.

The scale of this problem is well-documented in the broader telecommunications industry, though precise figures vary by methodology and market segment. The TM Forum's revenue assurance benchmarking work has reported leakage in the range of 1 to 5 percent of gross revenue [1]; its 2017/18 survey of operators reported an estimated average leakage rate of 1.9 percent, with individual responses reaching 5 percent or higher depending on measurement methodology [5]. Industry practitioners and consultancies have separately described revenue assurance as a strategic imperative, emphasizing that the underlying causes of leakage—billing system fragmentation, reconciliation gaps, and process complexity—remain persistent across operators [6]. Broadband-specific figures are less commonly published, but the structural causes—multi-system fragmentation, identifier inconsistency, and asynchronous event processing—are the same.

### 1.3 Reporting Friction and Audit Readiness

Operational reporting in many ISP environments still depends on manually maintained queries, ad hoc spreadsheet exports, and tribal knowledge about which source to trust for which metric. "Active subscriber count" might mean one thing to the finance team and something different to the network operations team. That kind of inconsistency is more than inconvenient. It undermines audit readiness, complicates the regulatory data submissions now required under FCC Broadband Data Collection rules [7, 8], and slows decisions that depend on accurate, timely numbers. Federal broadband deployment assessments have tracked these data quality challenges at the national level [9]. ISPs receiving federal broadband funding under the BEAD program face additional reporting obligations, including semi-annual infrastructure and subscriber data submissions, that further raise the stakes for data quality and consistency [10].

### 1.4 Positioning and Related Work

This paper does not attempt to survey the full landscape of enterprise data architecture, revenue assurance, or financial reporting compliance frameworks. The TM Forum has published extensive guidance on revenue assurance practices, including detailed reconciliation frameworks and maturity models [1, 11]. Cloud data platform vendors have published reference architectures for telecom analytics, and the TM Forum's Modern Data Architecture project has addressed the broader challenge of harmonizing OSS/BSS data across service providers [12]. In the financial reporting domain, the PCAOB, AICPA, and major audit firms have published extensive guidance on internal controls over financial reporting (ICFR), but their focus is on audit methodology rather than data engineering implementation. On the data engineering side, the multi-layer ELT pattern used here has been in common use since the early 2020s, most visibly as the "medallion architecture" popularized by Databricks [2], and general-purpose data engineering references now treat it as a common practice [13]. What is missing, in the publicly available literature reviewed for this paper, is an integrated approach. Existing tools tend to handle reconciliation, anomaly detection, semantic governance, and security controls as separate concerns. To the author's knowledge, no published practitioner methodology combines them into a single governed architecture. That apparent gap—between individual capabilities and an integrated framework—is what the GERA Framework is intended to address.

What is less well-documented is how to make these patterns work together. In a SOX-regulated environment, reconciliation without governance is just a matching exercise; governance without reconciliation leaves the underlying data problems untouched. This paper documents one approach to that integration: the GERA Framework, a vendor-neutral architecture that wires reconciliation, anomaly detection, semantic governance, and security controls into a single pipeline. The broadband case study provides the detail; the architecture itself has been carried across banking and enterprise finance environments by the author.

### 1.5 Intended Audience

This paper is written for data and platform engineering teams at regulated enterprises who are building or modernizing their analytics infrastructure. The primary case study is broadband, but the four-layer architecture applies wherever cross-system reconciliation, semantic governance, and audit-readiness matter—banking, enterprise finance, any SOX-regulated organization. Finance and compliance leaders evaluating the business case for structured reconciliation may also find it useful. The patterns are described at an architectural level rather than as implementation recipes; readers should expect to adapt them to their own technology stacks and organizational contexts.

## 2. ARCHITECTURAL APPROACH

### 2.1 Why Regulated Enterprises Need a Different Approach

Generic data warehouse patterns do not map cleanly onto regulated enterprise operations where financial reporting integrity, audit compliance, and cross-system reconciliation are requirements. Several characteristics set these environments apart from typical analytics settings. Broadband operations show them clearly.

First, the data is operationally coupled in ways that typical enterprise warehouses do not need to handle. A single fiber construction project generates financial commitments (purchase orders, invoices), physical material movements (receiving, issuance, installation), service activation events, and eventually billing records. These events span multiple source systems, arrive at different cadences, and use different identifiers—yet they must ultimately reconcile to a single economic reality.

Second, multi-party workflows are the norm, not the exception. Contractors receive materials, perform installations, and report completions through processes that may involve emailed spreadsheets, portal uploads, or batch file transfers. The variability in format, frequency, and quality is substantial, and most general-purpose ELT patterns underestimate it.

Third, the operational stakes are immediate. A billing mismatch does not just produce a bad report—it becomes a customer dispute. An undetected inventory surplus does not just skew a dashboard—it is trapped capital that could have funded the next phase of network buildout.

### 2.2 The GERA Framework: ELT Pipeline Architecture

The framework follows a modular Extract-Load-Transform (ELT) pattern, chosen deliberately over traditional ETL. Raw data is loaded into a cloud data warehouse first and transformed afterward, preserving the original records for auditability—a property that matters wherever financial reconciliation and audit trail preservation are requirements, which in practice means most SOX-regulated environments.

The framework organizes data into four layers, following the multi-layer pattern that has become common practice in modern data platforms—often called the "medallion architecture" in the data lakehouse community [2]. The GERA adaptation tailors it to regulated enterprise requirements: cross-system reconciliation, SOX-compliant audit trails, and governed semantic standardization. The broadband case study walks through these layers in detail. The author has deployed the same four-layer structure in a regulated banking environment and a Fortune 500 enterprise finance context. Figure 1 provides a high-level view.

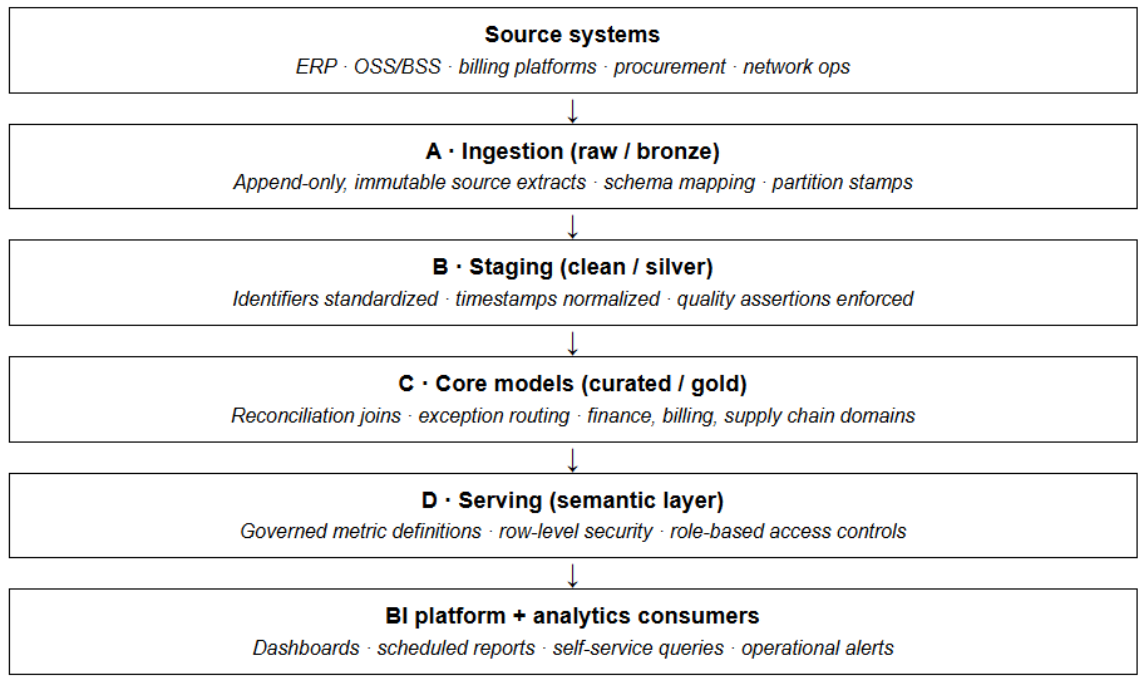

*Figure 1.* GERA Framework four-layer architecture (broadband case study). Raw source extracts enter an immutable ingestion tier (Layer A), pass through a staging layer for standardization (Layer B), feed into domain-specific core models with deterministic reconciliation logic (Layer C), and are served through a governed semantic layer with NIST CSF 2.0-aligned access controls (Layer D). Each tier has a single responsibility.

**Layer A: Ingestion (Raw/Bronze).** Source extracts and event payloads are stored in their original form. Only minimal normalization is applied—schema mapping, basic type casting, and partition stamping. The raw layer is append-only and immutable, providing a durable record that can be replayed if downstream logic changes.

**Layer B: Staging (Clean/Silver).** Identifiers, timestamps, and categorical fields are standardized. Deterministic cleanup rules handle the cross-system data quality problems that show up everywhere: inconsistent date formats across partner reports, leading zeros in material or account codes, variant spellings of entity names. In broadband, these are contractor report inconsistencies and location naming variants. In banking, they are GL account format discrepancies and cost center hierarchy mismatches. Data quality assertions (NOT NULL checks, accepted value ranges, referential integrity) catch problems early rather than letting them propagate [22].

**Layer C: Core Models (Curated/Gold).** Domain-specific models are assembled for finance, supply chain, billing and payments, and service operations—applying the principle of domain-oriented data ownership to model organization [14], though within a centralized platform rather than a fully decentralized data mesh. This is where reconciliation logic lives— cross-system joins that match provisioning events to billing records, purchase orders to receiving confirmations, issuance documents to installation reports. Unmatched records are routed to governed exception tables for investigation rather than silently dropped.

**Layer D: Serving (Semantic and Consumption).** A Semantic Layer governs metric definitions, separating business logic from visualization. When a dashboard shows "Active Subscriber Count," the definition traces back to a single governed calculation rather than a per-report formula. Role-based access controls enforce data visibility boundaries through the Enterprise BI Platform.

As the layer descriptions make clear, the architecture is designed so that each tier has a single responsibility: raw preservation, standardization, business logic, or governed consumption. This separation can simplify debugging, support incremental adoption, and make it possible to replace any single layer without disrupting the others.

*2.3 Semantic Layer and Metric Governance*

Metric inconsistency is one of the quieter but more corrosive problems in regulated enterprises. When the finance team's numbers do not match the operations team's numbers, every cross-functional conversation starts with a reconciliation exercise instead of a business decision. In broadband, this looks like subscriber count discrepancies between billing and network operations. In banking, it is GL balance mismatches across sub-ledger systems. The symptom differs; the cause is the same: no single governed definition.

The Semantic Layer addresses this by defining key metrics once, in a governed location, with explicit logic and source lineage. In a broadband context, governed metrics typically include active subscriber count (with clear rules for trial, pending, and suspended states), cost-per-passing and construction unit economics, inventory on-hand by location and aging classification, and billing reconciliation rates with exception categorization.

The concept of a semantic layer is not new—it has existed in enterprise BI since at least the dimensional modeling era [15]. However, the modern semantic layer has evolved significantly beyond its origins. Tools such as dbt's metrics layer (powered by MetricFlow) now define metrics as version-controlled code rather than embedded BI formulas, enabling governed metric definitions that are portable across visualization tools and queryable via APIs [16]. Other tools in this space—including Cube and AtScale—pursue similar goals, though a detailed comparison is outside this paper's scope. The Open Semantic Interchange (OSI) specification became publicly available in a GitHub repository as of January 27, 2026, representing an emerging industry effort toward vendor-neutral semantic model interchange [17]. Applying these capabilities specifically to the cross-domain data problems of broadband operations—where the same metric must reconcile billing, provisioning, and network data—is where the practical value lies.

*2.4 Cross-System Reconciliation Pattern*

Cross-system reconciliation failures typically originate not from fraud but from systems falling out of synchronization. In broadband, a service is activated in the provisioning system on Tuesday; the billing system picks it up on Thursday; the first invoice is generated on the next billing cycle. In banking, a general ledger entry posts in the ERP but the corresponding sub-ledger record lags or maps to a different cost center. Either way, if any step in this chain fails silently, the discrepancy may persist for months—and when it surfaces, it does so as an audit finding, a SOX deficiency, or a financial restatement. The TM Forum has documented this class of problem extensively in telecommunications [1, 11]. The underlying reconciliation pattern is the same regardless of industry [25].

The reconciliation pattern uses deterministic key matching to align events across systems. Deterministic matching—where records are joined on exact key values rather than probabilistic similarity scores—was chosen for this context because it produces transparent, auditable results and avoids the false-positive risk that probabilistic methods introduce in environments where billing accuracy is paramount. The tradeoff is that deterministic matching requires well-maintained identifier mappings and degrades when source identifiers are inconsistent; the constraints discussed in Section 4.3 address this directly.

The logic is straightforward in concept: define stable join keys that link a provisioning event to its corresponding billing record and payment settlement; maintain a governed exception table that captures every unmatched or inconsistent record;

track reconciliation status over time so that aging exceptions surface automatically rather than requiring someone to go looking.

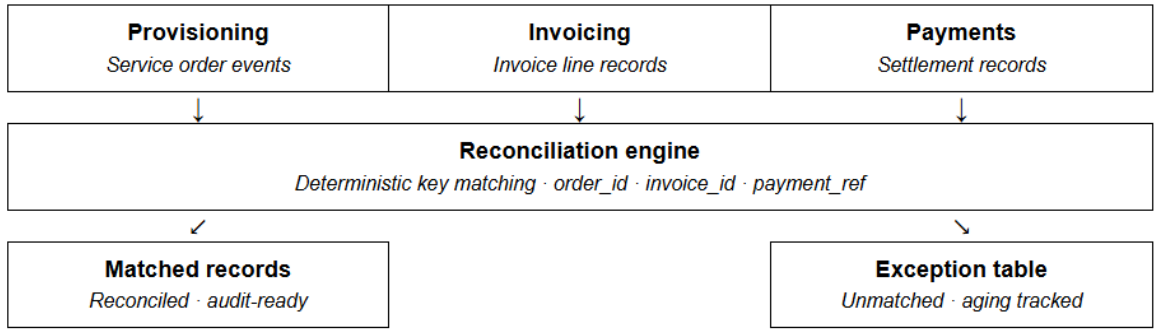

*Figure 2.* Billing reconciliation workflow. Provisioning events, invoice records, and payment settlements are joined on deterministic keys (order_id, invoice_id, payment_ref). Unmatched or inconsistent records are routed to a governed exception table rather than silently dropped—ensuring that mismatches become visible, trackable investigation items instead of silent data loss.

The harder part is handling the real-world messiness. Customer identifiers may not match exactly across systems. Timestamps may reflect different time zones or business-day conventions. Credits and adjustments may reference original transactions using inconsistent formats. A reconciliation model that assumes clean inputs will break quickly in production. The workflow in Figure 2 accounts for this by isolating exceptions rather than suppressing them.

**2.4.1 Illustrative Scenario: Service Activated but Not Billed**

To illustrate how the reconciliation pattern works in practice, consider a generalized scenario that reflects a common failure mode in broadband operations.

A residential fiber subscriber completes an online order for a 1 Gbps service plan. The order management system records the sale and passes it to the provisioning platform, which activates the optical network terminal and assigns a circuit identifier. From the subscriber's perspective, internet service is live within hours.

However, the provisioning event uses a ten-digit circuit identifier, while the billing platform expects a twelve-digit account number prefixed with a region code. An automated mapping between the two systems—maintained in a legacy middleware layer—fails silently for a subset of activations when a recent platform update changes the identifier format. No error is logged. The subscriber receives service; no invoice is generated.

Without a reconciliation layer, this mismatch goes undetected. The provisioning system shows an active subscriber. The billing system has no corresponding record. Neither system flags a discrepancy because neither is designed to check the other.

Under the reconciliation pattern described in this paper, a Core Model join attempts to match each provisioning event to a billing record using the mapped account key within a configurable time window—typically one billing cycle. When no match is found, the record is routed to a governed exception table with a status of "unmatched" and a timestamp marking the start of its aging clock. If the exception remains unresolved past a defined threshold—say, 14 days—it is escalated in the operational dashboard with an aging flag.

The operational value is straightforward: a revenue gap that might otherwise persist for weeks or months becomes a visible, trackable, and assignable investigation item within days of the activation event. The pattern does not determine root cause—that requires human investigation—but it is designed to help surface the problem before it compounds.

This type of scenario illustrates the general logic; the specific identifiers, thresholds, and system behaviors will vary by implementation.

*2.5 Illustrative Reconciliation Entities*

To make these patterns concrete, Table 1 shows a simplified example of how reconciliation entities might be structured in a broadband context. These are generalized illustrations, not implementation specifications. In practice, the clean join keys shown here are the goal, not the starting condition—the cross-system key challenges discussed in Section 4.3 apply directly.

*Table 1.* Illustrative GERA Reconciliation Entities (Broadband Case Study)

| Domain | Entity | Grain | Reconciliation Link |
|---|---|---|---|
| Billing | Service Order | Per activation | Join order_id + service_date to provisioning. Flag if no invoice within one billing cycle. |
| Finance | Invoice Line | Per charge item | Join invoice_id + subscriber_id to payment ledger. Flag if unpaid past settlement window. |
| Payments | Settlement | Per payment | Match payment_ref + invoice_id to invoice. Flag if orphaned payment with no source invoice. |
| Supply Chain | Issuance | Per job site | Join material_code + PO to receiving. Flag if issued but not installed past threshold. |
| Network Ops | Activation | Per circuit | Match circuit_id + activation_date to billing. Flag if active circuit with no billing record. |

In practice, these entities are joined through the Core Model layer (Layer C), with unmatched records routed to exception tables. Downstream dashboards then surface exception aging, resolution rates, and open dispute volumes—giving operations teams a structured view of where reconciliation is breaking down.

## 3. SECURITY AND GOVERNANCE

*3.1 NIST CSF 2.0–Aligned Controls*

Regulated enterprises face increasing scrutiny around data handling—not only for customer PII, but for financial reporting integrity, SOX compliance, and operational audit trails [3]. The governance layer described here is informed by the NIST Cybersecurity Framework (CSF) 2.0, mapping specific architectural controls to CSF 2.0 core functions. The examples below are drawn from broadband, but NIST CSF 2.0 is industry-agnostic by design—these governance patterns work the same way in a bank's data warehouse as they do in a telecom provider's. They do not constitute a complete compliance implementation; actual compliance depends on organizational policies, implementation rigor, and ongoing assessment [24].

CSF 2.0 organizes cybersecurity risk management into six functions: Govern, Identify, Protect, Detect, Respond, and Recover. The Govern function, new in CSF 2.0, is an overarching function focused on organizational governance strategy—establishing risk management priorities, defining roles and responsibilities, and integrating cybersecurity into enterprise decision-making (subcategories GV.RM, GV.RR, GV.PO) [3]. It is not a technical control but a management framework that informs all other functions. The data engineering patterns in this paper touch three functions at the technical implementation level:

**PROTECT — Access Control and Data Security (PR.AA, PR.DS):** Row-Level Security (RLS) provides dynamic filtering that can restrict data visibility by role and territory. A regional operations manager sees inventory and billing data for their service area only. Data access policies are defined declaratively through Infrastructure as Code tooling, so that access rules are version-controlled, reviewable, and reproducible across environments. This pattern supports the principle of least privilege (PR.AA) and data confidentiality (PR.DS-01), though its effectiveness depends on accurate role assignments and regular access reviews.

**DETECT — Continuous Monitoring (DE.CM):** Data access events are captured in an append-only audit log with defined retention periods. This log can support compliance reviews and incident investigations, provided it is monitored actively rather than treated as a passive archive. The pattern aligns with continuous monitoring of personnel activity and technology usage (DE.CM-03) and adverse event analysis (DE.AE).

**GOVERN — Organizational Context (GV.PO, GV.OV):** While the technical controls above implement protection and detection, their effectiveness depends on the organizational governance that CSF 2.0's Govern function emphasizes: documented policies, defined oversight responsibilities, and regular review cycles. Infrastructure as Code supports this by making governance decisions visible and auditable, but the organizational commitment to maintain and review those controls is a Govern-function responsibility, not a technical one.

*3.2 Infrastructure as Code*

The key principle is straightforward: governance controls should be reproducible. Access roles and permission groups are defined as code artifacts. Row-level filtering rules are deployed through the same change management process as application code. Permission changes produce an auditable history. This does not eliminate governance risk—no technical control does—but it can make governance decisions visible and reversible in a way that manual configuration cannot.

*3.3 Ethical and Privacy Considerations*

The patterns described in this paper process operational and financial data that may include customer identifiers, billing records, and service addresses. Although this paper does not prescribe a specific privacy implementation, the architectural design assumes a privacy-by-design posture: no personally identifiable information (PII) appears in the illustrative examples, all data access is governed by role-based controls aligned with the principle of least privilege (PR.AA), and audit logging supports accountability for data access events [3]. NIST's privacy engineering guidance provides a complementary risk management framework for organizations that need to address privacy risks alongside cybersecurity risks in data-intensive operational environments [18]. Organizations adopting these patterns should ensure that their implementations comply with applicable data protection requirements and that customer data handling aligns with their own privacy policies and regulatory obligations.

## 4. OPERATIONAL APPLICATIONS

*4.1 Inventory Optimization and Capital Efficiency*

Fiber deployment is capital-intensive, and inventory management is where capital efficiency is often won or lost. The failure mode is familiar to anyone who has walked a broadband warehouse: shelves of materials that have aged past usefulness, while procurement teams order replacements because the system of record does not reflect what is actually on the ground.

The inventory patterns address this through three mechanisms:

**Snapshot Generation.** Daily snapshots capture inventory state (snapshot_date, material_id, location_id, quantity_on_hand) to support historical trend analysis and point-in-time auditing. Without consistent snapshots, aging calculations lack a reliable baseline—and without a baseline, there is no way to distinguish normal fluctuation from a problem that needs attention.

**FIFO Aging Logic.** First-In-First-Out aging assigns each unit of inventory to a time bucket (0–30 days, 31–60 days, 61–90 days, >90 days) based on when it was received. Materials that have aged beyond a configurable threshold are surfaced for review. This is not a sophisticated algorithm, but it can be useful: in environments where aging visibility does not exist at all, even basic FIFO bucketing may surface significant trapped capital.

**Statistical Anomaly Detection.** A statistical layer flags inventory records where the observed quantity deviates significantly from the rolling mean. Statistical process control methods have been applied to anomaly detection in time-series data across domains [23]; the baseline implementation here uses a standard Z-Score calculation:

$$|z| = |(x - \mu) / \sigma| > 3$$

Records exceeding this threshold are flagged for physical audit. The intent is supportive, not authoritative—a high Z-score suggests a record warrants investigation, not that the data is necessarily wrong. Causes might include a legitimate bulk receipt, a data entry error, or a genuine discrepancy between the system and what is sitting on the shelf.

**Limitations and robust alternatives.** The standard Z-Score is presented here as a baseline method chosen for its simplicity and interpretability in environments where no anomaly detection currently exists. However, the statistical literature identifies significant limitations with this approach that practitioners should evaluate before production deployment. The Z-Score method assumes that inventory quantities are approximately normally distributed and uses the mean and standard deviation as its baseline statistics. In practice, inventory data is often non-normal—intermittent demand, seasonal patterns, and zero-inflated distributions are common in broadband materials management. When the underlying distribution is skewed, the standard Z-Score threshold of |z| > 3 may not correspond to meaningful probability levels, producing either excessive false positives or missed anomalies [19]. The mean and standard deviation are also sensitive to the very outliers the method is trying to detect: a single large discrepancy can inflate the standard deviation, masking subsequent anomalies. For these reasons, the robust statistics literature recommends the Modified Z-Score using Median Absolute Deviation (MAD)—which replaces the mean with the median and is substantially more resistant to outlier contamination—as the preferred default for outlier detection in non-normal data [20]. IQR-based (Interquartile Range) detection, which relies entirely on quantile statistics and makes no distributional assumptions, is another viable alternative. Organizations with inventory data that exhibits skewness, zero-inflation, or heavy tails should consider adopting MAD-based detection as their primary method rather than treating the standard Z-Score as sufficient.

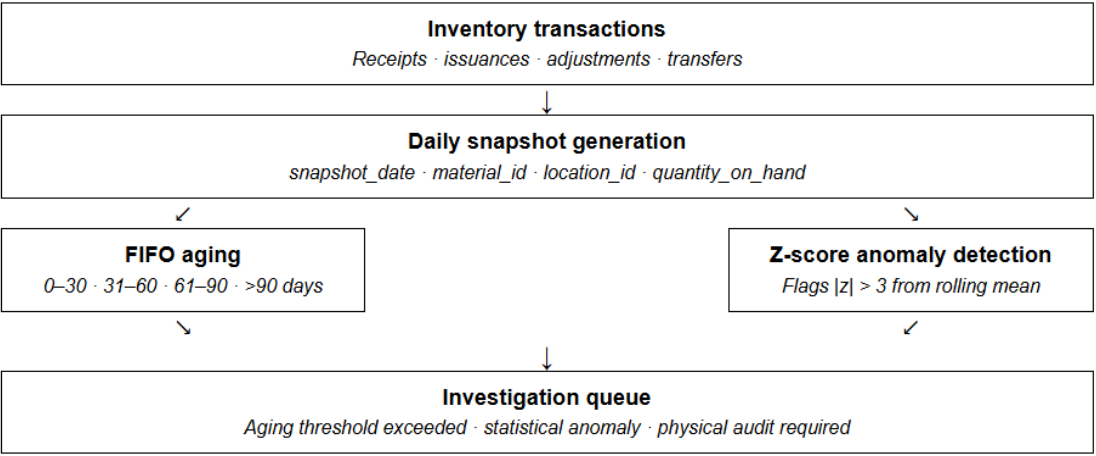

*Figure 3.* Inventory aging and anomaly detection workflow. Daily snapshots feed into FIFO aging buckets, while a parallel statistical calculation flags outliers. Records exceeding configurable thresholds are routed to an investigation queue for physical audit and disposition review. The two paths provide complementary coverage: FIFO catches materials that have simply sat too long; anomaly detection catches sudden quantity deviations that may indicate data entry errors or unreported movements.

**Data Fan-Out Mitigation.** In multi-party supply chains, a single purchase order can generate multiple receiving records, issuance documents, and installation confirmations. Without careful grain management, joins across these entities produce row multiplication that inflates totals and corrupts reports. The pattern mitigates this through pre-aggregation at stable grains before joining, deterministic deduplication, uniqueness constraints at intermediate stages, and isolation of exception records from core fact tables.

*4.2 Expected Operational Mechanisms*

These patterns target three categories of operational friction, though the magnitude of benefit depends heavily on the specific environment and starting condition:

**Manual reconciliation effort.** Automated cross-system matching can reduce the time analysts spend on manual data cleaning, exception investigation, and report preparation. Organizations starting from spreadsheet-based reconciliation workflows may see the most significant reduction in manual effort; those with existing automated processes are likely to see incremental improvement. The patterns do not eliminate manual investigation—they are designed to focus it on genuine exceptions rather than routine matching.

**Inventory carrying costs.** Aging analysis and anomaly detection can provide visibility into inventory that may be candidates for write-down, return, or reallocation. The value depends on the extent of existing aging visibility: organizations with no systematic aging analysis may discover that a material portion of on-hand inventory has aged beyond useful life, while those with mature inventory processes may find fewer surprises.

**Reporting and decision latency.** When governed metric definitions replace ad hoc per-report calculations, cross-functional discussions can proceed from shared numbers rather than preliminary reconciliation. This is an indirect benefit whose magnitude will vary by organization.

This paper makes no original quantitative performance claims; however, industry benchmarks provide a useful directional anchor. TM Forum survey data places average telecommunications revenue leakage at approximately 1.9 percent of gross revenue, with individual operators reporting materially higher rates [5], and the PCAOB reports a 39 percent aggregate Part I.A deficiency rate across all inspected firms [26]. Architectures that surface reconciliation exceptions earlier, standardize metric definitions, and strengthen audit trails are designed to address the failure modes associated with these categories of loss and control weakness, though this paper does not measure their effect. The patterns described here draw on practical experience across banking, telecommunications, and enterprise finance, but have not been validated through controlled measurement across independent implementations. Organizations evaluating adoption should assess expected benefit against their own operational baseline.

*4.3 Technical Constraints and Implementation Challenges*

A portable set of data engineering patterns must contend with real-world constraints that significantly affect implementation outcomes. The broadband case study surfaces them sharply, but they are just as present in banking and enterprise finance. The box below summarizes key considerations; the remainder of this section discusses each in more detail.

> **Implementation Considerations at a Glance**
> 1. **Late-arriving data** — Pipelines must be idempotent and support controlled look-back windows.
> 2. **Schema drift** — Ingestion layers need resilient schema handling and strong validation.
> 3. **Key mapping governance** — Cross-system identifier mappings require ongoing maintenance and version control.
> 4. **Grain control** — Pre-aggregate at stable grains before joining to prevent row multiplication.
> 5. **Exception handling ownership** — Assign organizational accountability for exception queue resolution.

**Late-arriving data** is common. In broadband, contractor reports and partner extracts may lag by days or weeks. In banking, month-end journal entries and sub-ledger adjustments arrive after the initial close. Pipelines must be idempotent and support controlled look-back windows to incorporate late records without duplicating earlier results.

**Schema drift** is persistent. Upstream systems change field names, add columns, or alter data types—often without advance notice to the data team. Ingestion layers need resilient schema handling and strong validation to prevent silent breakage.

**Cross-system key ambiguity** may be the most stubborn problem in this domain. Different systems use different customer identifiers, different material codes, different location hierarchies. The reconciliation approach described in this paper uses deterministic matching on canonical identifiers—but reaching that canonical form requires upstream normalization and crosswalk enrichment (e.g., standardized service IDs, mapped legacy account numbers, unified location codes). Records that cannot be deterministically matched after normalization are routed to a controlled exception workflow for manual review or constrained secondary matching, rather than silently dropped. This staged approach—normalize first, match deterministically on canonical keys, and quarantine non-matches—preserves the auditability benefits of deterministic methods while acknowledging that raw source identifiers are often inconsistent, partial, or evolving [21].

**Multi-party data variability** means that partner inputs arrive in different formats, at different frequencies, and at different quality levels. These patterns can help manage that variability, but they do not eliminate it.

**Cost and performance tradeoffs** require attention at scale. Incremental processing, intelligent partitioning, and careful grain selection are essential to keep pipeline runtimes and storage costs manageable as data volumes grow [13].

## 5. SCOPE BOUNDARIES AND LIMITATIONS

This paper documents the GERA Framework's data engineering patterns and architectural approaches. The broadband case study provides the detail; the patterns themselves apply to any regulated environment where cross-system reconciliation and governed reporting are requirements. It is a practitioner reference, not an empirical study, and does not claim more than that.

**These patterns do not replace systems of record.** The ELT architecture sits downstream of operational systems—ERP, OSS/BSS, billing, financial reporting, and procurement platforms. It consumes their data; it does not modify or govern it at the source. If source systems produce inaccurate data, downstream patterns can detect some inconsistencies but cannot correct them at their origin.

**Data quality outcomes depend heavily on identifier quality and source system discipline.** The reconciliation patterns are designed to work well when stable, deterministic join keys exist across systems. When identifiers are inconsistent, incomplete, or change over time, matching accuracy degrades. No amount of downstream engineering fully compensates for poor source data hygiene [21].

**Anomaly detection is supportive, not authoritative.** The statistical flagging described in the inventory management section identifies records that warrant investigation—it does not determine root cause or prescribe corrective action. False positives are expected and should be treated as part of a human-in-the-loop review process.

**Governance outcomes depend on implementation maturity.** Defining access policies as code and deploying Row-Level Security are necessary steps, but they are not sufficient by themselves. Sustained governance depends on ongoing access reviews, policy updates, monitoring, and organizational accountability that extend beyond the data engineering layer.

**This paper is not a compliance solution.** The NIST CSF 2.0 alignment described in Section 3 reflects architectural patterns that can support compliance goals, but actual regulatory compliance depends on organizational controls, policies, and audit practices that fall outside the scope of this work.

**Practitioner provenance.** The patterns are vendor-neutral by design. They draw on the author's direct implementation experience across three organizational contexts—a regional bank, a national broadband provider, and the central finance function of a Fortune 500 technology company—but they are presented as generalized reference approaches. No proprietary system configurations, internal datasets, or employer-specific implementation details are disclosed in this paper; the architectural patterns are abstracted to a level that is reproducible without access to any particular organization's infrastructure. The author has deployed the framework across these three contexts; it has not been independently replicated by other practitioners.

## 6. CONCLUSION

The cross-system reconciliation challenges facing regulated enterprises are persistent, and they resist generic solutions. The gap between an ERP transaction and a reporting record, the disconnect between a purchase order and a warehouse count, the inconsistency between one team's financial metric and another's—these are structural problems, not implementation bugs. They are rooted in how enterprise operations have evolved: systems acquired at different times, by different teams, for different purposes. The PCAOB's documented 39% aggregate Part I.A deficiency rate across all inspected firms confirms that these are not theoretical concerns [26].

The GERA Framework addresses these problems at the data engineering layer. Its contribution is integration: deterministic reconciliation that surfaces exceptions rather than hides them, statistical anomaly detection that flags records worth investigating, governed semantic standardization that defines organizational metrics once and traces them to source, and NIST CSF 2.0-aligned security controls that make access decisions reproducible. These capabilities exist individually in other tools. To the author's knowledge, no single published methodology in the literature reviewed for this paper wires them together in this form. The broadband case study provides the implementation detail; the same architecture has been deployed by the author in a regulated banking environment and a Fortune 500 enterprise finance context.

None of this replaces the operational judgment that enterprise teams bring to their work. The intent is to provide a more reliable data foundation underneath that judgment—so that experienced people spend their time on decisions rather than on reconciling numbers. Whether adopted in whole or adapted in parts, this is a practical starting point for regulated enterprises modernizing fragmented data environments. The broadband provider wrestling with billing reconciliation and the Fortune 500 finance team wrestling with SOX audit readiness face the same structural problem. The GERA Framework is one way to solve it.

## REFERENCES

[1] TM Forum (Business Assurance Project), "GB941 Main Revenue Assurance Guidebook," ver. 3.6.0, published Dec. 22, 2023, TM Forum Approved Feb. 9, 2024. [Online]. Available: https://www.tmforum.org/resources/guidebook/gb941-main-revenue-assurance-guidebook-v3-6-0/ (login required; accessed Apr. 6, 2026).

[2] Databricks Staff, "What is Medallion Architecture?" Databricks Blog, n.d. [Online]. Available: https://www.databricks.com/blog/what-is-medallion-architecture (accessed Apr. 6, 2026).

[3] National Institute of Standards and Technology, "The NIST Cybersecurity Framework (CSF) 2.0," NIST CSWP 29, Feb. 26, 2024. DOI: 10.6028/NIST.CSWP.29. [Online]. Available: https://doi.org/10.6028/NIST.CSWP.29 (accessed Apr. 6, 2026).

[4] U.S. Government Accountability Office, "Broadband Programs: Agencies Need to Further Improve Their Data Quality and Coordination Efforts," GAO-25-107207, Apr. 17, 2025 (publicly released Apr. 28, 2025). [Online]. Available: https://www.gao.gov/products/gao-25-107207 (accessed Apr. 6, 2026).

[5] TM Forum, "TM Forum Revenue Assurance Survey Report 2017/18 Edition," ver. 1.1, Mar. 29, 2018. [Online]. Available: https://www.tmforum.org/wp-content/uploads/2018/03/TMF_Revenue-Assurance_Survey_201718_v1_1.pdf (accessed Apr. 6, 2026).

[6] PwC, "Revenue Assurance: A Strategic Imperative in Today's Complex Business Landscape," Dec. 24, 2024. [Online]. Available: https://www.pwc.com/m1/en/publications/revenue-assurance-strategic-imperative-in-todays-complex-business-landscape.html (HTML) and https://www.pwc.com/m1/en/publications/documents/2024/pwc-revenue-assurance.pdf (PDF). Accessed: Apr. 6, 2026.

[7] Broadband Deployment Accuracy and Technological Availability Act ("Broadband DATA Act"), Pub. L. No. 116–130, 134 Stat. 228, Mar. 23, 2020. [Online]. Available: https://www.govinfo.gov/app/details/PLAW-116publ130 (accessed Apr. 6, 2026).

[8] Federal Communications Commission, "Establishing the Digital Opportunity Data Collection; Modernizing the FCC Form 477 Data Program," Third Report and Order, FCC 21-20, adopted Jan. 13, 2021, released Jan. 19, 2021. [Online]. Available: https://docs.fcc.gov/public/attachments/fcc-21-20a1.pdf (accessed Apr. 6, 2026).

[9] Federal Communications Commission, "2024 Section 706 Report," FCC 24-27, GN Docket No. 22-270, adopted Mar. 14, 2024, released Mar. 18, 2024. [Online]. Available: https://docs.fcc.gov/public/attachments/FCC-24-27A1.pdf (accessed Apr. 6, 2026).

[10] National Telecommunications and Information Administration, "Semi-Annual Performance (Technical) Report v2.0 Form," BEAD Program, due Jan. 30, 2026, n.d. [Online]. Available: https://broadbandusa.ntia.gov/sites/default/files/2026-01/NTIA_BEAD_Semi-Annual_Technical_Report_v2.0_Form_01_26.pdf (accessed Apr. 6, 2026).

[11] TM Forum (Revenue Management Project), "TR131 Revenue Assurance Overview," ver. 2.4.1, team approved Apr. 13, 2012; date modified May 6, 2014 (archived; superseded by v2.5.0). [Online]. Available: https://www.tmforum.org/resources/technical-report/tr131-revenue-assurance-overview-v2-4-1/ (login required; accessed Apr. 6, 2026).

[12] TM Forum (Modern Data Architecture project), "IG1356 Data Architecture for AI-enabled Telecom Operations Whitepaper," ver. 2.0.0, team approved Mar. 1, 2024; published May 6, 2024; modified Jun. 10, 2024 (archived; superseded by v3.0.0). [Online]. Available: https://www.tmforum.org/resources/introductory-guide/ig1356-data-architecture-for-ai-enabled-telecom-operations-whitepaper-v2-0-0/ (login required; accessed Apr. 6, 2026).

[13] J. Reis and M. Housley, Fundamentals of Data Engineering: Plan and Build Robust Data Systems. Sebastopol, CA, USA: O'Reilly Media, Jun. 2022. ISBN: 9781098108304. [Online]. Available: O'Reilly or Google Books (accessed Apr. 6, 2026).

[14] Z. Dehghani, Data Mesh: Delivering Data-Driven Value at Scale. Sebastopol, CA, USA: O'Reilly Media, 2022. ISBN: 9781492092391. [Online]. Available: Google Books, https://books.google.com/books/about/Data_Mesh.html?id=M5J5zgEACAAJ (accessed Apr. 6, 2026).

[15] R. Kimball and M. Ross, The Data Warehouse Toolkit: The Definitive Guide to Dimensional Modeling, 3rd ed. John Wiley & Sons, Jul. 2013. ISBN: 9781118530801. [Online]. Available: https://www.wiley.com/en-us/The+Data+Warehouse+Toolkit%2C+3rd+Edition-p-9781118530801 (accessed Apr. 6, 2026).

[16] dbt Labs, "dbt Semantic Layer," dbt Developer Hub, last updated Apr. 2, 2026. [Online]. Available: https://docs.getdbt.com/docs/use-dbt-semantic-layer/dbt-sl (accessed Apr. 6, 2026).

[17] Snowflake, "Open Semantic Interchange (OSI) Updates: Specification Now Live," Snowflake Blog, Jan. 27, 2026. [Online]. Available: https://www.snowflake.com/en/blog/open-semantic-interchanges-specs-finalized/ Also: OSI spec repository (Apache-2.0), https://github.com/open-semantic-interchange/OSI (accessed Apr. 6, 2026).

[18] National Institute of Standards and Technology, "NISTIR 8062: An Introduction to Privacy Engineering and Risk Management in Federal Systems," Jan. 2017. [Online]. Available: https://nvlpubs.nist.gov/nistpubs/ir/2017/nist.ir.8062.pdf (accessed Apr. 6, 2026).

[19] B. Iglewicz and D. C. Hoaglin, How to Detect and Handle Outliers, vol. 16, ASQC Basic References in Quality Control. Milwaukee, WI, USA: ASQC Quality Press, 1993 (print edition). ISBN: 087389247X (9780873892476). [Online]. Available: WorldCat or HathiTrust (accessed Apr. 6, 2026).

[20] C. Leys, C. Ley, O. Klein, P. Bernard, and L. Licata, "Detecting Outliers: Do Not Use Standard Deviation Around the Mean, Use Absolute Deviation Around the Median," J. Exp. Soc. Psychol., vol. 49, no. 4, pp. 764–766, Jul. 2013. [Online]. DOI: 10.1016/j.jesp.2013.03.013. Available: https://doi.org/10.1016/j.jesp.2013.03.013 (accessed Apr. 6, 2026).

[21] DAMA International, DAMA-DMBOK: Data Management Body of Knowledge, 2nd ed. Rev. Bradley Beach, NJ, USA: Technics Publications, 2017. ISBN: 9781634622349. [Online]. Available: Technics Publications (accessed Apr. 6, 2026).

[22] M. Souibgui, F. Atigui, S. Zammali, S. Cherfi, and S. Ben Yahia, "Data quality in ETL process: A preliminary study," Procedia Computer Science, vol. 159, pp. 676–687, 2019. DOI: 10.1016/j.procs.2019.09.223.

[23] D. Zheng, F. Li, and T. Zhao, "Self-adaptive statistical process control for anomaly detection in time series," Expert Systems with Applications, vol. 57, pp. 324–336, 2016. DOI: 10.1016/j.eswa.2016.03.029.

[24] P. Mikalef, M. Boura, G. Lekakos, and J. Krogstie, "The role of information governance in big data analytics driven innovation," Information & Management, vol. 57, no. 7, p. 103361, Nov. 2020. DOI: 10.1016/j.im.2020.103361.

[25] R. Álvarez-Foronda, C. De-Pablos-Heredero, and J.-L. Rodríguez-Sánchez, "Implementation model of data analytics as a tool for improving internal audit processes," Frontiers in Psychology, vol. 14, p. 1140972, 2023. DOI: 10.3389/fpsyg.2023.1140972.

[26] Public Company Accounting Oversight Board, "Spotlight: Staff Update on 2024 Inspection Activities," Mar. 31, 2025. [Online]. Available: https://pcaobus.org/documents/staff-update-2024-inspection-activities-spotlight.pdf (accessed Apr. 15, 2026).